# Enhanced tunability of two-dimensional electron gas on SrTiO$_3$ through heterostructuring

*Hyang Keun Yoo[1,2,3], Luca Moreschini[1], Aaron Bostwick[1], Andrew L. Walter[4], Tae Won Noh[2,3], Eli Rotenberg[1], and Young Jun Chang[5,]\**

1. Advanced Light Source, E. O. Lawrence Berkeley National Laboratory, Berkeley, California 94720, USA

2. Center for Correlated Electron Systems, Institute for Basic Science (IBS), Seoul 08826, Republic of Korea

3. Department of Physics and Astronomy, Seoul National University, Seoul 08826, Republic of Korea

4. Photon Sciences Directorate, Brookhaven National Laboratory, NSLS II, Upton, New York 11973, USA

5. Department of Physics, University of Seoul, Seoul, 02504, Republic of Korea

*E-mail: yjchang@uos.ac.kr

**Abstract**

Two-dimensional electron gases (2DEGs) on the $SrTiO_3$ (STO) surface or in STO-based heterostructures have exhibited many intriguing phenomena, which are strongly dependent on the 2DEG-carrier density. We report that the tunability of the 2DEG-carrier density is significantly enhanced by adding a monolayer $LaTiO_3$ (LTO) onto the STO. Ultraviolet (UV) irradiation induced maximum carrier density of the 2DEG in LTO/STO is increased by a factor of ~4 times, compared to that of the bare STO. By oxygen gas exposure, it becomes 10 times smaller than that of the bare STO. This enhanced tunability is attributed to the drastic surface property change of a polar LTO layer by UV irradiation and $O_2$ exposure. This indicates that the 2DEG controllability in LTO/STO is more reliable than that on the bare STO driven by defects, such an oxygen vacancy.

# Introduction

The two-dimensional electron gas (2DEG) is one of the most interesting systems in condensed matter[1-6]. The 2DEG has been formed at the interface of semiconductor heterostructure, which reveals many novel phenomena including integer/fractional quantum hall effects with extremely high mobile carriers[4-6]. As a result, the 2DEG system has also been investigated for applications to the high mobility electronic devices[7,8]. The 2DEG in transition-metal oxides could exhibit more various novel properties due to interplay between spin/charge/orbital degrees of freedoms[9,10]. Furthermore, since the 2DEG can be artificially created at the interface of the transition-metal oxide heterostructure[2,11-14], such emerging phenomena will be applicable for next generation electronic devices.

The 2DEG on $SrTiO_3$ (STO) has recently attracted much attention due to many fascinating properties, including high mobility[15], giant thermoelectric Seebeck coefficient[16,17] and 2D superconductivity[18]. Additionally, the 2DEG exhibits more intriguing phenomena near the interface of the heterostructure. For example, the $LaAlO_3$/STO heterostructure shows an exotic 2D superconductivity, such as coexistence with ferromagnetism, at the interface[19,20]. A new high-$T_c$ superconductor has been recently discovered in the monolayer FeSe on STO heterostructure[21-23]. These interesting behaviors in the STO or STO based heterostructures are strongly dependent on the 2DEG-carrier density (CD). Therefore, the controllability of the 2DEG-CD could be one of the key ingredients to understand and manipulate these novel phenomena. The simplest way is *in-vacuum* annealing by varying the oxygen partial pressure[24]. The amount of the oxygen vacancies directly controls the CD. However, this method is usually difficult to apply in precise control due to complex interplay between control parameters, such as vacuum level, temperature, annealing and cooling durations. Thus, the development of a new simple method is highly required.

Here, by using angle-resolved photoemission spectroscopy (ARPES), we investigated the tunability of 2DEG through synchrotron ultraviolet (UV)-irradiation and molecular oxygen gas ($O_2(g)$)-exposure. The 2DEG-CD of bare STO increases with UV-irradiation and decreases with $O_2(g)$-exposure. However, the tunability of the 2DEG-CD is significantly enhanced by deposition of a monolayer $LaTiO_3$ (LTO) on the STO. The maximum 2DEG-CD in LTO/STO is increased by a factor of 4 times under UV-irradiation, compared to that of the bare STO. Additionally, with $O_2(g)$-exposure, it becomes 10 times smaller than that of the bare STO. This enhanced tunability is attributed to the drastic surface property change of a polar LTO layer, compared to that of nonpolar STO.

**Results and Discussion**

On the bare STO surface, the 2DEG is created by UV-irradiation. First, we prepared an *in-vacuum* annealed (1000°C, $10^{-9}$ Torr, 5 min) STO (001) single crystal[25]. Then, as shown in Fig. 1(a), we exposed the bare STO to UV intensity ($I_{UV}$) of 3.2 W cm$^{-2}$ until the ARPES intensity near the Fermi level ($E_F$) became saturated. As shown in Fig. 1(b), we observed at least two electron-like bands with concentric circular shapes. The two concentric Fermi surfaces correspond to the $d_{xy}$ orbital bands, which has two-dimensional character [26,27] Fermi wave numbers ($k_F$) of the shallower and deeper bands are 0.123 and 0.176 Å$^{-1}$, respectively. And the former and latter bands have band bottoms estimated to be at 0.12 and 0.2 eV, respectively. We draw the white dotted lines as estimated band dispersions, by roughly matching the simple parabolic dispersion curves to the intense features near the Fermi level[25-27]. We calculated the surface CD by extracting the Luttinger area from the Fermi surface map. The total surface CD is around $7 \times 10^{13}$ cm$^{-2}$. These results are consistent with previous reports[26,27]. The proposed model for the formation of 2DEG is surface band bending with UV-irradiation induced oxygen desorption, and charge carrier accumulation near the surface[26-29]. We also note that there are weak intensities along the $k_x$ and $k_y$ axes (black arrows in Fig. 1(b)), which are probably due to the previously observed shallow elliptic Fermi surfaces (the $d_{yz}$ or $d_{xz}$ orbital bands)[26].

By exposing a bare STO to $O_2(g)$, we can reduce the CD significantly. As shown in Fig. 1(c), we exposed the bare STO to $O_2(g)$ for 3 langmuir, i.e. 5 minutes at $1 \times 10^{-8}$ Torr. During the exposure process, $I_{UV}$ is set to 0.05 W cm$^{-2}$ to minimize the UV-dose effect[27]. In Fig. 1(d), we observed only one electron band which has $k_F$ = 0.128 Å$^{-1}$ and band bottom = 0.12 eV. The estimated CD is around $3.5 \times 10^{13}$ cm$^{-2}$. This result indicates that the UV-induced oxygen desorption near the STO surface is partially compensated by the $O_2(g)$-exposure. As a result, the surface electrons from the oxygen vacancies are reduced, which makes the reductions of surface band bending and charge carrier accumulation[26-29]. The amount of observed CD change is about a factor of 2, rather small for device applications. Note that, the previous report of the 2DEG control in fractured STO surface shows more significant response with $O_2(g)$-exposure[29], compared to that of our polished STO surface. We expected that it originates from more concentrated oxygen vacancy formation near surface, not inside bulk, in the fractured case. However, origins for the different properties between fractured and polished STO surfaces are still unclear, so further investigations are demanded.

We found that, upon depositing a monolayer LTO on STO surface, the tunability of 2DEG becomes significantly enhanced. As shown in Fig. 1(e), we prepared a monolayer LTO on STO (001) substrate by using pulsed laser deposition (PLD) method. Note that the surface LTO layer is already exposed to a high concentration of $O_2(g)$ during PLD growth, so in order to achieve a high CD, we needed to illuminate UV light to maximize the occupation of the 2DEG state in LTO/STO. We exposed the LTO/STO sample to $I_{UV}$ = 3.2 W cm$^{-2}$ until the PES intensity near $E_F$ was saturated. In Fig. 1(f), we observed four electron-like bands, such as circular and ellipsoidal ones corresponding to $d_{xy}$ and $d_{xz,yz}$ orbitals respectively[30,31]. Because of strong matrix element effects, the intensity along the Fermi contours is highly variable[31], so that only small sections of each band are revealed in the figure. The dominant CD originates from circular $d_{xy}$ orbital. The shallower $d_{xy}$ band (with $k_F = 0.22$ Å$^{-1}$) and deeper $d_{xy}$ band (with $k_F = 0.36$ Å$^{-1}$) have band bottoms at around 0.2 and 0.4 eV, respectively. Although the rough parabolic curve fitting gives the band bottom, we could not analyze the energy dispersion curves along the long residue intensity, where quasi-particle interactions are previously studied[28,32,33]. The total surface CD in LTO/STO is around $3 \times 10^{14}$ cm$^{-2}$. This result indicates that the UV-exposure can generate 4 times larger 2DEG-CD in the LTO/STO than that in the bare STO. The observed band structure is consistent with previous results for 2DEG in a stoichiometric monolayer LTO/STO system[30,31]. Thus, we expect that the polar LaO$^+$ and TiO$_2^-$ layers can strengthen the band bending near the surface, which induces a larger 2DEG band and a higher charge carrier accumulation.

The 2DEG-CD in LTO/STO can be also changed by O$_2$(g)-exposure. Surprisingly, we found that the amount of change is much larger than that of bare STO case. As shown in Fig. 1(g), we exposed the LTO/STO sample with CD of $3 \times 10^{14}$ cm$^{-2}$ to O$_2$(g) for 3 langmuir. Then, we perform ARPES measurements with $I_{UV} = 0.05$ W cm$^{-2}$. In Fig. 1(h), we observed a very small one electron band which has $k_F = 0.04$ Å$^{-1}$ and band bottom = 0.05 eV. The estimated CD is around $4 \times 10^{12}$ cm$^{-2}$: namely, the O$_2$(g)-exposure on LTO/STO can suppress the CD by more than 70 times. This amount of CD change in LTO/STO is much larger than the factor 2 change, observed in the bare STO. This experimental result indicates that the O$_2$(g) compensation effect for 2DEG in the LTO/STO is also stronger than that in the bare STO.

To understand the difference between bare and LTO-covered STO systems, we measured the real-time Ti core-level photoemission spectra during O$_2$(g)-exposure. As shown in Fig. 2(a), first, we exposed the bare STO, having a maximum 2DEG-CD shown in Fig. 1(b), to O$_2$(g) in partial pressure, $1 \times 10^{-8}$ Torr. During O$_2$(g)-exposure, we used $I_{UV} = 0.05$ W cm$^{-2}$ to reduce the UV-dose effect. The Ti 2p$_{3/2}$ core-level is ~0.1 eV shifted to lower binding energy, indicating the reduction of the surface band bending, and saturated in the 2DEG state of the minimum

CD. To more precisely analyze the change of the core-level, we draw the spectra of initial (red line) and final (black dotted line) ones in Fig. 2(b). The initial Ti $2p_{3/2}$ core-level shows two peaks at 459.6 and 457.4 eV, corresponding to $Ti^{4+}$ and $Ti^{3+}$, respectively[34]. This $Ti^{3+}$ could be attributed to the oxygen vacancy states in the bare STO due to UV-induced oxygen desorption. However, after $O_2(g)$-exposure, the $Ti^{3+}$ peak becomes very small. This indicates that the oxygen desorption is readily compensated by $O_2(g)$, which induces a dominant $Ti^{4+}$ valence near the bare STO surface.

We also measured the Ti $2p_{3/2}$ core-level in LTO/STO sample with the same experimental condition of bare STO (Fig. 2(c)). We exposed the $O_2(g)$ to the LTO/STO, having a maximum 2DEG-CD shown in Fig. 1(d). The core-level is ~0.3 eV shifted to lower binding energy, indicating the reduction of surface band bending, and saturated to the minimum 2DEG-CD. As shown in Fig. 2(d), the initial Ti $2p_{3/2}$ core-level also shows two peaks corresponding to $Ti^{4+}$ and $Ti^{3+}$, in which the $Ti^{3+}$ peak is much larger than the bare STO case. The stoichiometric LTO layer has $Ti^{3+}$ state due to the valence state of $La^{3+}$, donating one more electron than $Sr^{2+}$. Considering the effective probing depth (1~2 nm) of these core-level photoemission measurements ($h\nu$ = 700 eV), it is roughly estimated to probe 3~5 unit cells of top layers, covering the top-most LTO layer and a few underneath STO layers[31]. This explains that the largely increased intensity of $Ti^{3+}$ peak is mostly due to the LTO layer and possibly some oxygen vacancy states in the underneath STO layer. Therefore, in Fig. 2(d), the $Ti^{3+}$ and $Ti^{4+}$ signals are largely originated from the stoichiometric surface LTO layer and the underneath STO layers, respectively. Note that, the valence band structure in Fig. 1(f) indicates the stoichiometric $La^{3+}Ti^{3+}O_3^{6-}$ layer in the case of the maximum 2DEG[30,31]. Then, after $O_2(g)$-exposure, $Ti^{3+}$ peak is almost erased. This indicates that the LTO layer experiences a large valence change, so that it could be changed into a new phase which has a $Ti^{4+}$ valence by $O_2(g)$-exposure. It is also important to note that such changes in the core-level states are reversible during the cycles of $O_2(g)$-exposure and UV-irradiation, which implies some sort of reversible oxygen interactions at the surfaces.

We speculated that the $O_2(g)$-exposure makes a $La^{3+}Ti^{4+}O_{3.5}^{7-}$ phase, showing an insulating behavior[35], but the UV-irradiation recovers the $La^{3+}Ti^{3+}O_3^{6-}$ phase on STO. Namely, the UV-irradiation induced stoichiometric $La^{3+}Ti^{3+}O_3^{6-}$ monolayer on STO can have a larger 2DEG and higher CD than that of the bare STO due to surface dipole formation[31]. On the other hand, with $O_2(g)$-exposure, the surface LTO layer becomes an insulating $La^{3+}Ti^{4+}O_{3.5}^{7-}$ phase which is energetically stable in high $O_2(g)$ pressure[36]. As a result, the charge carriers are transferred to the excess oxygen in $La^{3+}Ti^{4+}O_{3.5}^{7-}$ phase, which induces ideally no 2DEG state in LTO/STO. It is worth noting $LaTiO_{3+\delta}$ ($\delta \neq 0.5$) phase shows a metallic behavior[35]. Thus, if the $O_2(g)$-exposure makes a majority $LaTiO_{3+\delta}$ ($\delta \neq 0.5$) phase, we expected that the minimum 2DEG size would be comparable with that of the bare $SrTiO_{3-\delta}$. However, we observed a small 2DEG state in the experiment after $O_2(g)$-exposure. It is still unclear whether the observed 2DEG state arises from a minority $La^{3+}Ti^{3+}O_3^{6-}$ phase with $O_2(g)$-exposure or from extrinsic defects formation, such as oxygen vacancies, during film deposition or other origins. In addition, it is of great importance to reveal surface morphology change during the $O_2(g)$-exposure and UV-irradiation. We note that atomic-scale interaction between oxygen molecules and surface structures should occurs depending on gas pressure and surface reconstructions in a reversible manner[37,38]. Thus, further experimental and theoretical investigations are highly required by employing surface structure analysis tools, such as scanning tunneling microscopy or surface x-ray scattering[38,39].

To get further information, we performed a real-time work function ($W_F$) measurement during UV-irradiation ($I_{UV}$ = 2.2 W cm$^{-2}$) first without, and then with $O_2(g)$-exposure ($1 \times 10^{-8}$ Torr). We found that, as shown in Fig. 3(a), the $W_F$ is almost the same in the bare STO in spite of changes in the surface band bending and 2DEG-CD. In semiconductors, the $W_F$ is kept to be nearly constant although the charge carriers are redistributed near the surface or the interface[40]. This means that the control of the 2DEG on the bare STO can be explained by the conventional semiconductor physics with a surface doping change. Contrary to the case of bare STO, $W_F$ of 2DEG in LTO/STO is altered by both UV- and O$_2$(g)-exposure. As shown in Fig. 3(b), the $W_F$ value of LTO/STO, which has a ~0.3 eV higher value than that of the bare STO, decreases with UV-irradiation, whereas it increases with O$_2$(g)-exposure. The amount of $W_F$ change is ~0.1 eV. This result indicates that the change of 2DEG in LTO/STO cannot be explained by the conventional semiconductor physics. Instead, our proposal of the surface LTO phase change can explain this intriguing $W_F$ behavior in LTO/STO. The reported $W_F$ of the La$^{3+}$Ti$^{4+}$O$_{3.5}$$^{7-}$ is 4.45 eV[41], consistent with our result. As the insulating La$^{3+}$Ti$^{4+}$O$_{3.5}$$^{7-}$ layer becomes polar TiO$_2$$^-$/LaO$^+$ layer with UV-irradiation, the $W_F$ can become lower due to the surface dipole formation[42].

We further investigated the tunable property of 2DEG in LTO/STO. As shown in Fig. 4(a), we schematically illustrated that the intense UV light induces the 2DEG states on the designated area and then oxygen gas exposure suppresses the 2DEG states on the entire surface of LTO/STO. In Fig. 4(b), we wrote the letter 'ALS' with the intense UV beam ($I_{UV}$ = 4.6 W cm$^{-2}$) and recorded the 2DEG intensity with much weaker UV intensity $I_{UV}$ = 0.05 W cm$^{-2}$. Then, we gradually exposed the same sample with O$_2$(g) (0.5 and 1.5 langmuir) and recorded the gradually disappearing 2DEG intensity. After 1.5 langmuir exposure, the sample surface is reset with minimal 2DEG signal, confirming erasing of letter 'ALS'. The data acquisition time in Fig. 4 is much shorter than those in Fig. 1(f,h), so that the signals of 2DEG in Fig. 4 are weaker than those in Fig. 1(f,h). This cycle of writing and erasing have been repeated on the several samples to demonstrate the reversible and reproducible behavior of 2DEG states in the LTO/STO heterostructure. This result suggests a new platform to investigate the nano-size heterojunction properties, such as memristive behavior at the interface of metal/2DEG[43] and charge transfer behavior at the interface of the exfoliated 2D materials, i.e. graphene and transition-metal dichalcogenides[44], on 2DEG by using nano-size UV beam patterning and *in situ* sample preparation methods[45-47]. Finally, based on the reversible tunability, the LTO/STO can be used as a new material for the nano-size $O_2(g)$ sensor since the patterned 2DEG on LTO/STO is very sensitive to the $O_2(g)$[47,48]. This sensor can be repeatedly used by UV patterning after $O_2(g)$ sensing.

**Conclusion**

In summary, we found that the tunability of 2DEG through UV-irradiation and $O_2(g)$-exposure is significantly enhanced by adding a monolayer LTO on STO. We suggested that this enhanced tunability originates from the drastic surface property change of polar LTO layer in both processes, compared to that of nonpolar STO. We demonstrated one example that heterostructuring could enhance the tunability of the novel functionalities for oxide electronics applications. We expected that our finding would accelerate the searching for enhanced tunable functionality through heterostructuring in other strongly correlated transition metal oxides.

**Methods**

*Sample preparation*: High-quality, single crystals of STO were used for this experiment. All the atomically flat STO substrates were prepared using buffered-HF (BHF) etching. For the 2DEG formation on the bare STO, the BHF-treated STO single crystal is annealed *in-vacuum*, i.e. 1000°C, $10^{-9}$ Torr, 5 min. It was exposed to UV of 3.2 W cm$^{-2}$ until the intensity near $E_F$ is saturated, which indicates the maximum size of 2DEG on bare STO. When the $O_2(g)$ was exposed for 3 langmuir, the minimum size of 2DEG was achieved. On the other hand, high-quality epitaxial monolayer LTO/STO film was fabricated using pulsed laser deposition. The details of fabrication method are described in ref. 31. Note that, the surface of LTO is already exposed to the high concentration of $O_2(g)$ during deposition, so we needed to UV dose to maximize occupation of the 2DEG state in LTO/STO. We exposed the LTO/STO sample with $I_{UV}$ = 3.2 W cm$^{-2}$ for the maximum size of 2DEG. Then, as the $O_2(g)$ was exposed for 3 langmuir, the minimum size of 2DEG was achieved.

*In situ Photoemission spectroscopy*: All measurements were performed at an end-station equipped with ARPES and PLD, sharing the same ultrahigh vacuum envelope, at the Beamline 7.0.1 of the Advanced Light Source. After sample preparation, such as *in-vacuum* annealing of bare STO and deposition of a monolayer LTO on STO, the samples were transferred without breaking a vacuum to an analysis chamber with a base pressure of $5 \times 10^{-11}$ Torr. The sample temperature was in 100~150 K. ARPES measurements were carried out in the Γ plane at $h\nu$ = 95 eV, which was already surveyed by varying the photon energy at normal emission. The total-energy resolution was about 25 meV at 95 eV. The measurement power density of photon source was kept at 0.05 W cm$^{-2}$, except for UV-dose cases, to minimize the UV dosing effect on the sample surfaces. On the other hand, the work function $W_F$ was determined in the ARPES instrument using the secondary electron cut-off.

**Acknowledgements**

This work was supported by the Basic Study and Interdisciplinary R&D Foundation Fund of the University of Seoul (2019) for Y.J. Chang. The Advanced Light Source is supported by the Director, Office of Science, Office of Basic Energy Sciences, of the U.S. Department of Energy under Contract No. DE-AC02-05CH11231. This work was supported by IBS-R009-D1.

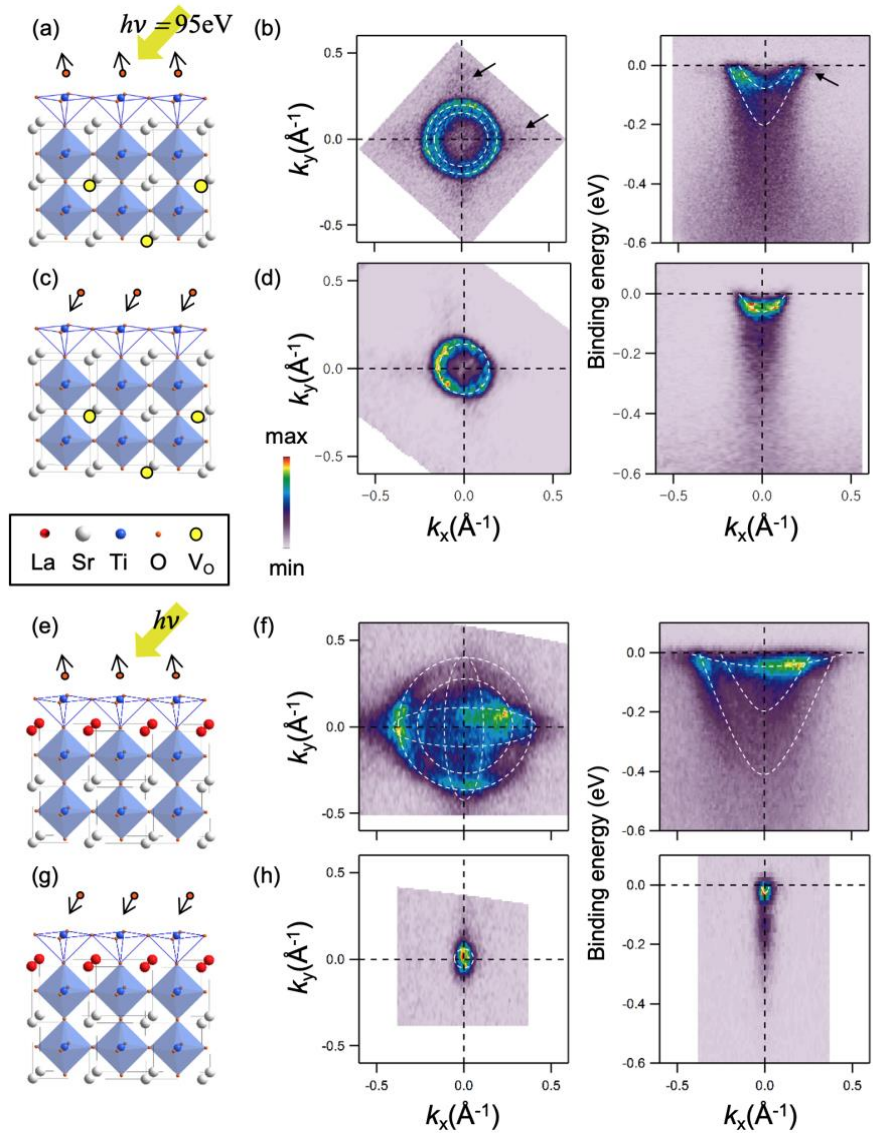

**Figure 1.** Observation of a tunable behavior of 2DEG through UV-irradiation and $O_2(g)$-exposure. (a)-(d) Schematic diagram of tuning the surface 2DEG on the bare STO, and the corresponding Fermi surface map (left) and ARPES spectra (right) after (b) UV-irradiation and (d) $O_2(g)$-exposure. Note that the 2DEGs in (b) and (d) are maximum and minimum sizes, respectively. The same experiments are performed in the monolayer LTO on STO sample, shown in (e)-(h). The maximum and minimum sizes of 2DEGs in LTO/STO become larger and smaller than those of the bare STO, respectively. This result indicates that the tunability of the 2DEG is enhanced by adding the monolayer LTO on STO. Note that the white dotted lines are guides to the eye

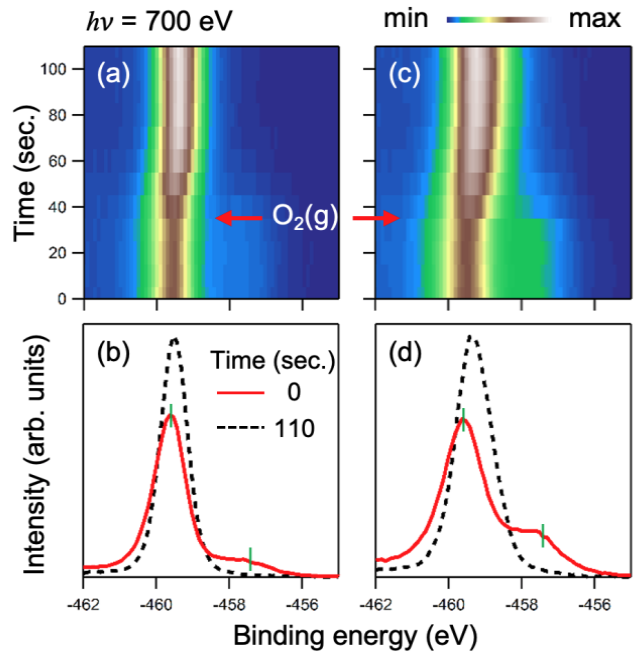

**Figure 2.** Real-time Ti core-level photoemission spectra during $O_2(g)$-exposure. (a)-(d) Experimental results for the bare STO and the monolayer LTO on STO. The spectra of initial and final ones are shown in (b), (d) by red and black-dotted lines, respectively. The initial Ti $2p_{3/2}$ core-level spectra show two peaks, corresponding to $Ti^{4+}$ and $Ti^{3+}$, respectively, marked by green markers. However, after $O_2(g)$-exposure, $Ti^{3+}$ peak becomes very small in both cases.

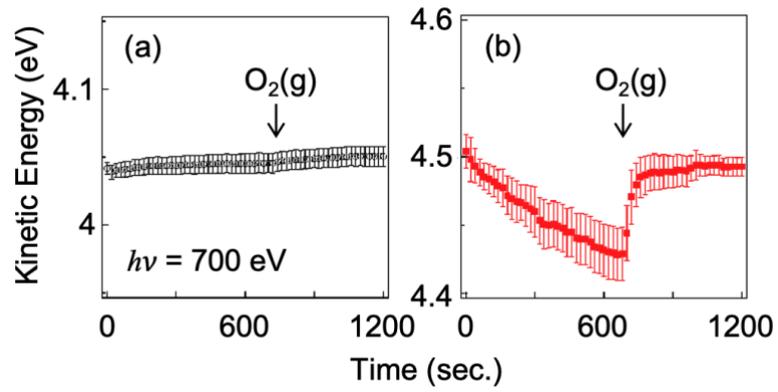

**Figure 3.** Real-time $W_F$ measurement during UV-irradiation and $O_2(g)$-exposure. The $W_F$ of the bare STO and the LTO/STO are shown in (a) and (b), respectively. The bare STO reveals a negligible change during UV-irradiation and $O_2(g)$-exposure, whereas the LTO/STO exhibit a significant change during both processes.

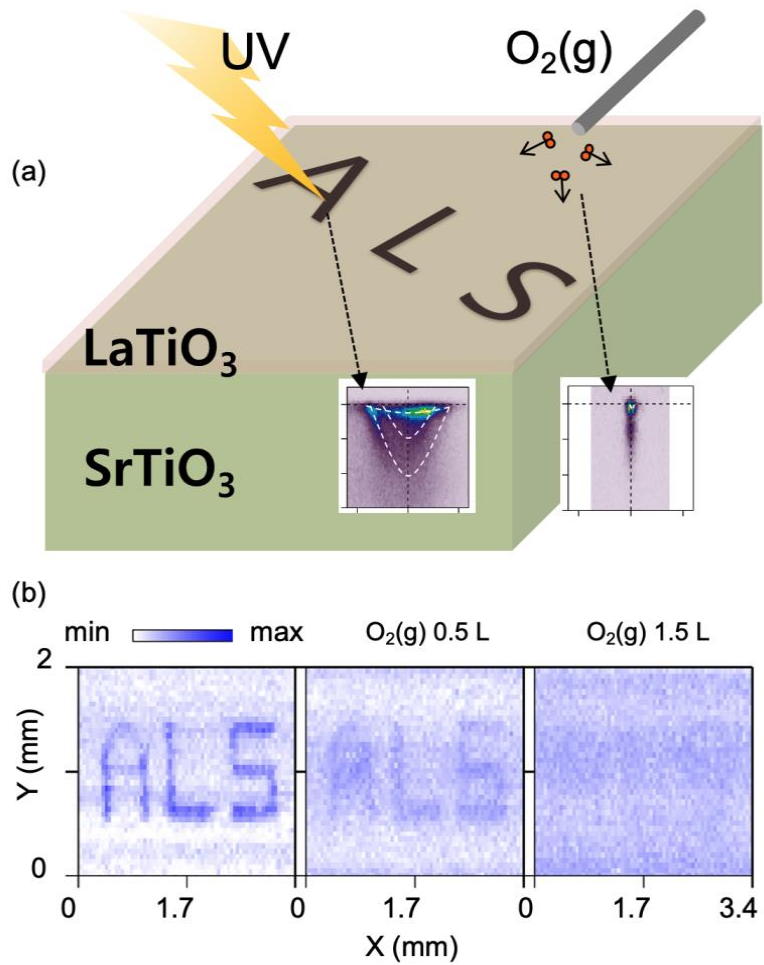

**Figure 4.** (a) The schematic drawing of the 2DEG state patterning by UV beam and O$_2$(g) exposure. The UV beam generates surface 2DEG state on the selective area and the following oxygen gas exposure erases the 2DEG state. (b) Experimental demonstration of the patterning of the abbreviation of Advanced Light Source, ALS, on LTO/STO. The intense signal corresponds to the ARPES spectra intensity of 2DEG states in Fig. 1(f) after the UV beam irradiation. When exposed to oxygen gas, the 2DEG intensity weakens in 0.5 langmuir and disappears in 1.5 langmuir. This cycle of writing and erasing is repeated to demonstrates reversible character of such UV induced oxygen desorption and oxygen exposure on this LTO/STO surface.